\documentclass[journal,oneside,final]{IEEEtran}
\usepackage[T1]{fontenc}
\usepackage{amsthm}
\usepackage{amsbsy}
\usepackage{amstext}
\usepackage{graphicx}
\usepackage{fixltx2e}
\usepackage{stackengine}
\usepackage{amssymb}
\usepackage{url}
\usepackage{float}
\usepackage{color}
\usepackage{epstopdf}
\usepackage{amsmath}
\usepackage{cite}
\usepackage{balance}
\usepackage{lastpage}
\usepackage{flushend}
\usepackage{lipsum}
\usepackage{caption}
\usepackage{subcaption}
\newcommand{\bi}{\begin{itemize}}
\newcommand{\ei}{\end{itemize}}

\newcommand{\be}{\begin{enumerate}}
\newcommand{\ee}{\end{enumerate}}

\newcommand{\ed}{\end{description}}
\newcommand{\bc}{\begin{center}}
\newcommand{\ec}{\end{center}}
\newcommand{\bt}{\begin{tabbing}}
\newcommand{\et}{\end{tabbing}}
\newcommand{\bfig}{\begin{figure}}
\newcommand{\efig}{\end{figure}}
\newcommand{\beq}{\begin{equation}}
\newcommand{\beqarr}{\begin{eqnarray}}
\newcommand{\beqarrn}{\begin{eqnarray*}}
\newcommand{\eeq}{\end{equation}}
\newcommand{\eeqarr}{\end{eqnarray}}
\newcommand{\eeqarrn}{\end{eqnarray*}}
\newcommand{\bflr}{\begin{flushright}\vspace{-0.2in}}
\newcommand{\eflr}{\end{flushright}}
\newcommand{\bsub}{\begin{subequations}}
\newcommand{\esub}{\end{subequations}}
\newcommand{\barr}{\begin{array}}
\newcommand{\earr}{\end{array}}

\begin{document}
\title{\huge On Anomalous Diffusion of Devices in Molecular Communication Network}
%
\vspace{-.5cm}
\author{Lokendra Chouhan, Prabhat Kumar Upadhyay, Prabhat Kumar Sharma, and  Anas M. Salhab
\thanks{L. Chouhan and P. K. Upadhyay are with the Department of Electrical Engineering, IIT Indore, India (e-mail: lokendrachouhan22@gmail.com, pkupadhyay@iiti.ac.in).
}\thanks{P. K. Sharma is with the Department of Electronics and Communication Engineering, VNIT, Nagpur, India (e-mail: prabhatmadhavec1@gmail.com)
}
\thanks{A. M. Salhab is with the Department of Electrical Engineering, KFUPM, Saudi Arabia, (e-mail: salhab@kfupm.edu.sa)
}
\vspace{-1.4cm}}
\maketitle
\begin{abstract}
A one-dimensional (1-D) anomalous-diffusive molecular communication channel is considered, wherein the devices (transmitter (TX) and receiver (RX)) can move in either direction along the axis. For modeling the anomalous diffusion of information carrying molecules (ICM) as well as that of the TX and RX, the concept of time-scaled Brownian motion is explored.
In this context, a novel closed-form expression for the first hitting time density (FHTD) is derived. Further, the derived FHTD is validated through particle-based simulation.
For the transmission of binary information, the timing modulation is exploited.
Furthermore, the channel is assumed as a binary erasure channel (BEC) and analyzed in terms of achievable information rate (AIR). 
\end{abstract}
\begin{IEEEkeywords}
Anomalous diffusion, mobile molecular communication, scaled-Bownian motion.
\end{IEEEkeywords}
\section{Introduction}
Molecular communication (MC) is an interdisciplinary area of research that has gained the community's interest due to its promising applications in intracellular therapy, chrono drug delivery, tissue engineering, lab-on-chip, and toxic gas monitoring \cite{nakano2012molecular}. 

In the existing literature, MC channels are classified as classical-diffusive and anomalous-diffusive channels~\cite{cao2015anomalous}. In contrast to classical-diffusive channels, the mean square displacement (MSD) of information carrying molecules (ICM) scales varies non-linearly with time. The anomalous diffusion occurs in many applications, such as spatially disordered systems like plasmas and turbulent fluids. Moreover, anomalous diffusion can also be found in biological media with traps, receptor binding sites, or macromolecular crowding \cite{ryabov2003behavior}.

Several works in literature, including \cite{cao2015anomalous,7506289,7856998,8830431,trinh2020molecular}, have analyzed the anomalous-diffusive channel in MC systems. The first hitting time density (FHTD) for the anomalous-diffusive medium has been obtained in \cite{cao2015anomalous} using the Caputo fractional derivative, and the channel has been analyzed in terms of bit error rate.
In \cite{7506289}, the authors considered the super-diffusive channel and provided a technique for optimum detection at the receiver (RX).  An online event detection method which can cope with fractional diffusion was proposed in \cite{7856998}. The stochastic behavior of anomalous diffusive channel was analyzed in \cite{8830431}.
However, in \cite{trinh2020molecular}, H-diffusion model was used to analyze the anomalous diffusion phenomenon in MC using a timing modulation method.
The anomalous diffusion-based research have been classified in \cite{chouhan2020molecular}, wherein a one-dimensional (1-D) anomalous-diffusive channel was analyzed considering the scaled-Brownian motion of ICM. 

In \cite{cao2015anomalous,7506289,7856998}, and \cite{chouhan2020molecular}, fixed transmitter (TX) and receiver (RX) were considered in MC channel. However, in many practical scenarios, TX and RX can also diffuse anomalously with the ICM into the MC  channel, e.g., water molecules in brain tissue, turbulent plasma, bacterial motion, amorphous semiconductors, and the porous media \cite{metzler2000random}. In this context, in \cite{8830431}, the stochastic behaviour of anomalous-diffusive channel has been studied considering random positioning of TX and RX. The mobile anomalous-diffusive channel under random time constraints has been studied in \cite{trinh2020molecular}. Note that in \cite{8830431,trinh2020molecular}, a continuous time random walk (CTRW) model has been used, which shows the non-Gaussian behavior and mathematical intractability.

To the best of authors' knowledge, none of the earlier works has focused on the study of anomalous diffusion of TX and RX into the MC channel.  Nevertheless, the mobility of TX and RX needs to be considered and worth to be investigated inside the anomalous-diffusive channel, as it relates to many applications like disease detection and targeted drug delivery. In the existing literature, the widely studied MC channel is a binary symmetric MC channel. For example, in \cite{chouhan2020molecular,channelcapacity1} and references therein, a binary symmetric MC channel has been considered and analyzed in terms of channel capacity.
However, due to the anomalous diffusion of ICM as well as the devices such as transmitter (TX) and receiver (RX), the arrival probability of ICM in the intended time-slot is very less~\cite{Transposition,chouhan2020molecular}. Therefore, the possibility of ICM in out-of-order arrival at the RX is very high. Eventually, the channel may reaches into the erasure state. Thus, a MC system should be studied for the binary erasure channel (BEC) considering timing modulation scheme.
Based on the aforementioned motivations, following are the new contributions through this paper:
\begin{itemize}
\item A 1-D \emph{mobile} anomalous-diffusive channel, where TX and RX can also move along with the ICM, is explored.
\item In contrast to \cite{8830431,trinh2020molecular}, for anomalous diffusion phenomena, the concept of scaled Brownian motion (SBM) is used
and a novel closed-form expression for the FHTD is derived.
\item Moreover, the MC channel is considered as a BEC and analyzed in terms of achievable information rate (AIR) incorporating timing modulation for the binary transmission.
\item Further, the classical, super- and sub-diffusive channels are compared with each other on the basis of AIR and FHTD.
\end{itemize} The derived FHTD is validated using the stochastic particle-based simulation (SPBS) approach.
  \vspace{-0.3cm}
\section{Anomalous-Diffusive Channel Model}\label{sm}
We consider a 1-D anomalous-diffusive \emph{mobile} MC channel along the $z$-axis. Let the non-reacting ICM, TX and RX can move inside the MC channel with the diffusion coefficients $D_{m}$, $D_{tx}$ and $D_{rx}$, respectively.
We use the time-scaled Brownian motion concept to analytically model the anomalous diffusion of ICM, TX and RX, inside the MC channel. In case of time-scaled Brownian motion, the MSD of the ICM, TX and RX at any arbitrary time $t$, is defined as: MSD $\sim 2 D_{\phi} t^{\alpha}$, for $\phi\in\{m,tx,rx\}$, where $\alpha$ denotes the anomalous diffusion exponent. The subscripts \textit{m, tx} and \textit{rx} signify the notations for molecule, TX and RX, respectively.
Based on the value of anomalous-diffusion exponent $\alpha$ which ranges over $0 \leq \alpha \leq 2$, three classes of diffusion phenomena are defined as: (1) normal-diffusion ($\alpha = 1$), (2)  sub-diffusion ($\alpha\leq 1$), and (3) super-diffusion ($\alpha> 1$) \cite{chouhan2020molecular}.
The instantaneous diffusion coefficient $D_{\phi}(t)$ corresponding to the entity $\phi$ is time-varying and depends on effective diffusion coefficient ($D_{\phi}$) as
$D_{\phi}(t)= \alpha t^{\alpha-1}D_{\phi}$ \cite{chouhan2020molecular}.
%


\underline{\textbf{FHTD for \emph{Static} TX and RX:}} Let the TX and RX initially at $t=0$ placed along the $z$-axis on the positions $z_{0,tx}$ and $z_{0,rx}$, respectively. The initial distance ($r_{0}$) between the TX and RX is given as $r_{0} = \sqrt{{(z_{0,tx} - z_{0,rx})}^2}$.
Let the molecule be released from TX at time $T_r$ = $kT_s$, i.e., at the beginning of the $k$th sampling interval with duration $T_s$. The molecule arrival time at the RX is a random variable and given by $T_a = T_r +T_h$, where $T_h$ denotes the first hitting time of a molecule.
For the fixed TX and RX, which are placed at a given distance $r_0$ with each other inside the 1-D anomalous-diffusive channel, the FHTD is given as~\cite{chouhan2020molecular}
\vspace{-.19cm}
\begin{align}
f_{T_h}(t_h|r_0) = \frac{r_0}{\sqrt{4\pi D_m (t_h)^{\alpha+2}}}\exp\left(\frac{-(r_0)^2}{4D_m(t_h)^{\alpha}}\right),\label{e:3}
\end{align} where $t_h$ denotes a dummy first-hitting time variable.

\underline{\textbf{FHTD for \emph{Mobile} TX and RX:}} Due to the scaled Brownian motion of TX and RX, their positions in the 1-D anomalous channel are also random. Let the $z_{k,tx}$ and $z_{k,rx}$ be the random positions of TX and RX at the $k$th instance, respectively. The change in position $\Delta z_{\Phi}$ of entity $\Phi$ in sampling interval $T_s$ is defined as $\Delta z_{\Phi}\sim\mathcal{N}(0,2D_{\Phi}{(T_s)}^{\alpha})$~\cite{chouhan2020molecular}, where $\mathcal{N}(u,v)$ denotes the Gaussian random variable with  mean $u$ and variance $v$, and $\Phi\in\{tx,rx\}$. Now the position of the entity $\Phi$ at $k$th sampling interval is given as $z_{k,\Phi} = z_{0,\Phi}+\sum_{\nu =1}^{k}\Delta z_{\Phi}$,
where $z_{k,\Phi}$ follows the Gaussian distribution as $z_{k,\Phi}\sim\mathcal{N}(z_{0,\Phi},2kD_{\Phi}(T_s)^{\alpha})$.
\begin{figure}
  \centering
  \includegraphics[width=3.5in, height = 1.6in]{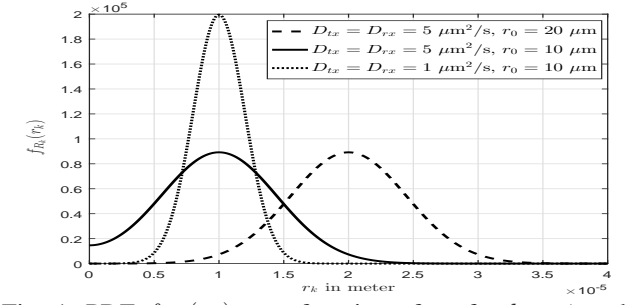}
  \vspace{-0.6cm}
  \caption{PDF $f_{R_k}(r_k)$ as a function of $r_{k}$ for $k =1$ and $T_s = 1$\,s (cf. \eqref{eqa:1A}).}
  \vspace{-0.7cm}\label{fig:1AA}
\end{figure}
\begin{figure*}[t]
\begin{subfigure}[b]{0.34\textwidth}
                \centering
                \includegraphics[width=\linewidth, height = 1.5in]{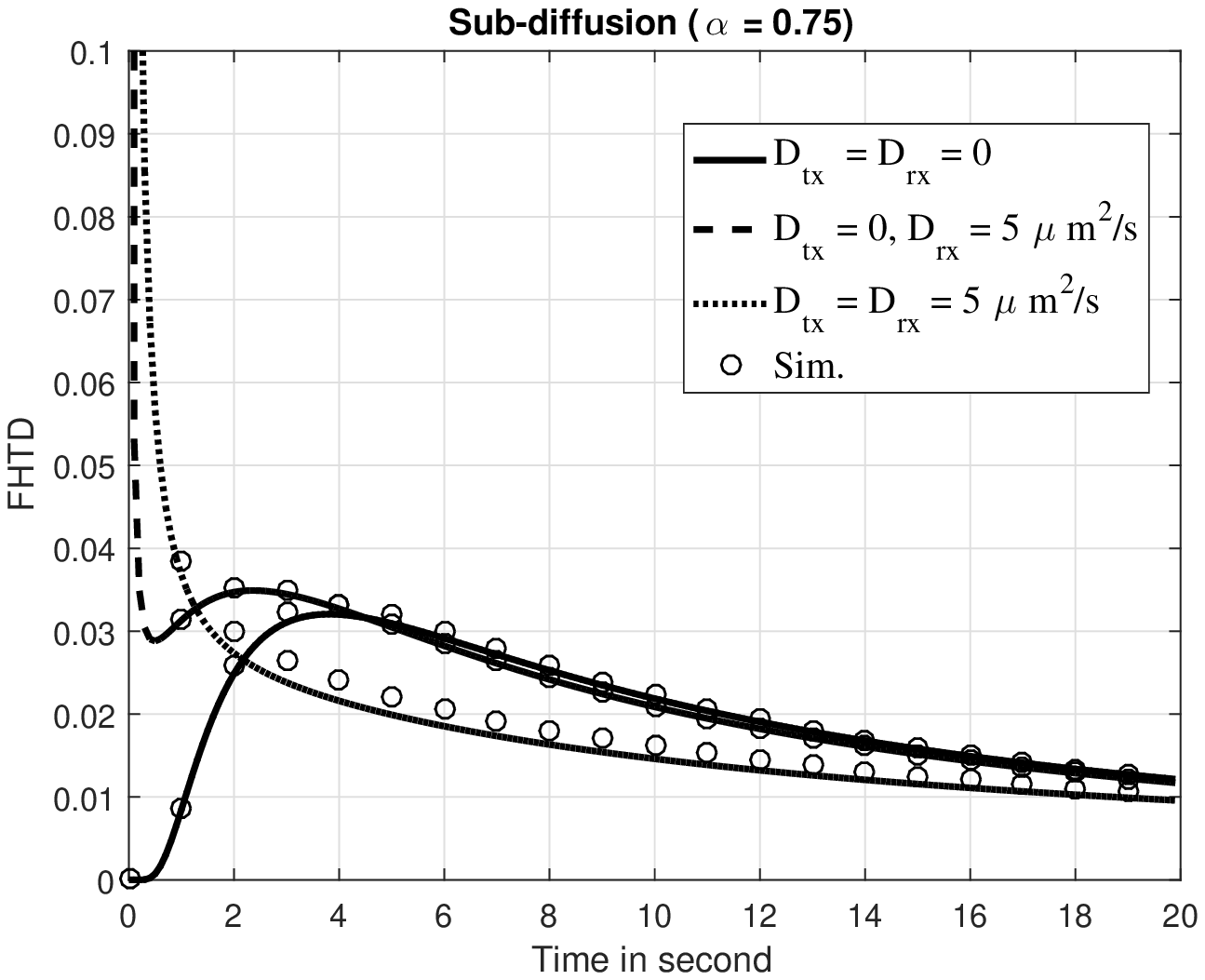}
                \label{fig3_a}
        \end{subfigure}%
        \begin{subfigure}[b]{0.34\textwidth}
                \centering
                \includegraphics[width=\linewidth, height = 1.5in]{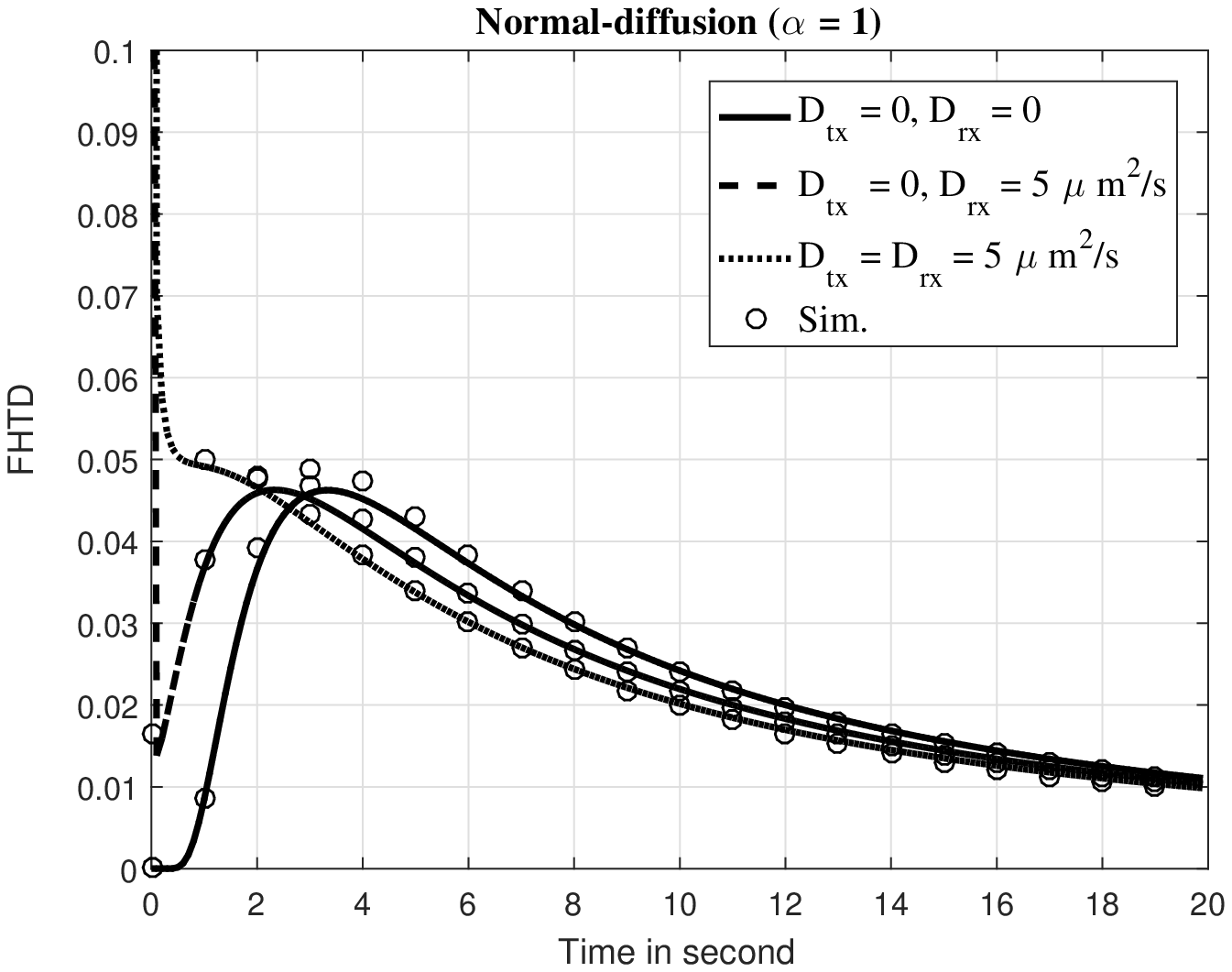}
                \label{fig3_b}
        \end{subfigure}%
        \begin{subfigure}[b]{0.34\textwidth}
                \centering
                \includegraphics[width=\linewidth, height = 1.5in]{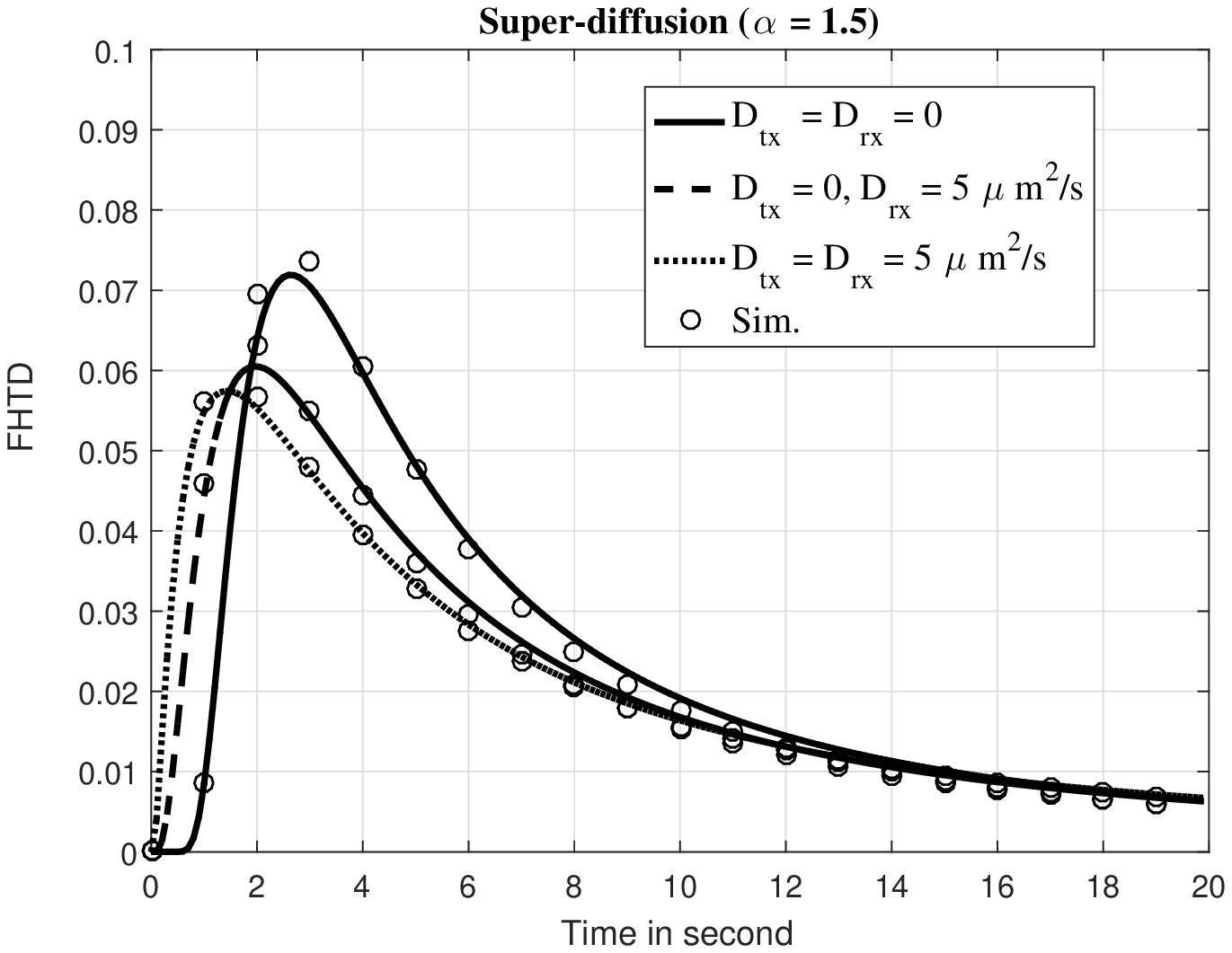}
                \label{fig3_c}
        \end{subfigure}%
        \vspace{-.5cm}
        \caption{FHTD as a function of time for different diffusion phenomena and different sets of $D_{tx}$ and $D_{rx}$.}\label{fig3}
        \vspace{-.6cm}
    \end{figure*}

Now the random distance between TX and RX at the $k$th sampling interval is defined as $r_k = \sqrt{{(z_{k,tx} - z_{k,rx})}^2}$. Using the property of subtraction of two Gaussian random variables, we have $(z_{k,tx} - z_{k,rx})\sim \mathcal{N}\left(\left(z_{0,tx}-z_{0,rx}\right),2kD_{t,r}{(T_s)}^{\alpha}\right)$, where $D_{t,r} = D_{tx}+D_{rx}$ known as effective diffusion coefficient capturing the relative diffusivity of TX and RX.

Let $f_{{R_k}}(r_k)$ be the PDF of the random distance $r_k$ between the mobile TX and RX in the $k$th sampling  interval.
The PDF $f_{{R_k}}(r_k)$ follows the non-central chi distribution \cite{Transposition}, and can be written as
\begin{align}
f_{{R_k}}(r_k) = &\frac{1}{\sqrt{2\pi\sigma_k^2}}\text{exp}\left(-\frac{r_0^2}{2\sigma_k^2}\right)\notag\\&\times\sum_{l=1}^2\exp\Bigg(-\frac{r_k^2}{2\sigma_k^2}+{(-1)}^{l}\frac{r_0 r_k}{\sigma_k^2}\Bigg),\label{eqa:1A}
\end{align}where $\sigma_k = \sqrt{2k{(T_s)}^{\alpha}D_{t,r}}$ is the standard deviation of the displacement between the mobile TX and RX
For the case when TX and RX are \emph{mobile}, we use the concept of relative diffusivity of ICM, i.e., the diffusion coefficient $D_{m}$ in \eqref{e:3} is replaced by the effective diffusion coefficient $D_{m,r} = D_{m}+D_{rx}$ \cite{PhysRevLett.82.1578}.
By averaging $f_{T_h}(t_h|r_k)$ over all the possible random distances between TX and RX, the FHTD for $k$th sampling interval can be obtained as
\begin{align}
\overline{f_{T_h}(t_h;k)} &= \int_{r_k = 0}^\infty f_{T_h}(t_h|r_k) f_{{R_k}}(r_k) dr_{k} \nonumber \\
&=\!\int_{r_k = 0}^\infty\!  \frac{r_k}{\sqrt{4\pi D_{m,r} (t_h)^{\alpha+2}}}\!\exp\!\left(\!\frac{-r_k^2}{4D_{m,r}(t_h)^{\alpha}}\!\right)\!  \nonumber \\
& \qquad \qquad\times f_{{R_k}}(r_k) dr_{k}. \label{eq:10AA}
\end{align}
Now using \eqref{eqa:1A}, the above integral can be expanded as follows
\begin{align}
&\overline{f_{T_h}(t_h;k)}  =  \int_{r_k = 0}^\infty \frac{r_k}{2\sigma_k\pi\sqrt{ 2D_{m,r}(t_h)^{\alpha+2}}}\text{exp}\!\left(-\frac{r_0^2}{2\sigma_k^2}\right)\notag\\ &\! \times \! \sum_{l=1}^2\exp\Bigg(\!\!-r_k^2\bigg(\!\frac{1}{4D_{m,r}(t_h)^{\alpha}}\!+\frac{1}{2\sigma_k^2}\!\bigg){+}{(-1)}^{l}\frac{r_0 r_k}{\sigma_k^2}\!\Bigg)\!dr_{k}.\nonumber
\end{align}
Now the integration above can be solved using \cite[Eq. (3.462-5)]{gradshteyn2014table}. The final expression for FHTD is obtained as
\begin{align}
 \overline{f_{T_h}(t_h;k)}&=\frac{ab e^{\frac{-r_0^2}{2b}}}{t_h\pi (a+b)}+\frac{r_0}{t_h}\sqrt{\frac{1}{4\pi{(a+b)}^3}}\notag\\ &\times \text{exp}\left({\frac{-r_{0}^2}{4(a+b)}}\right) \text{erf}\left(\frac{r_0}{2}\sqrt{\frac{a}{b(a+b)}} \right),\label{eq:18A}
\end{align}
where \text{erf}($\cdot$) denotes the error function, $a = D_{m,r} (t_h)^{\alpha}$, and  $b = k D_{t,r} {(T_s)}^{\alpha}$. Note that for fixed TX and RX, i.e., with $D_{tx} = D_{rx} = 0$, \eqref{eq:18A} reduces to \eqref{e:3}.

\begin{figure}
  \centering
  \includegraphics[width=3.5in, height = 1.8in]{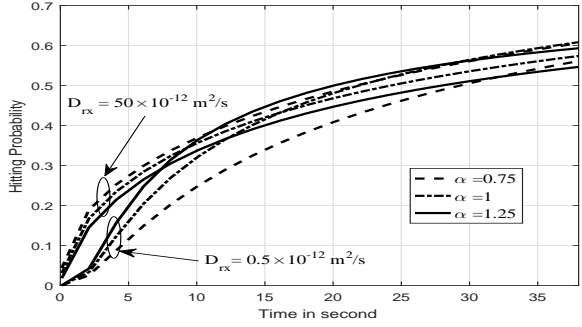}
  \vspace{-0.6cm}
  \caption{Hitting probability as a function of time.}
  \vspace{-0.6cm}\label{fig:3}
\end{figure}

Since both TX and RX are mobile in a 1-D medium at the molecule release time $T_r = kT_s$, there is a non-zero probability that the distance between TX and RX is 0.
Further in simulation, we consider that after the collision, the TX and RX reflect back to their previous positions~\cite{cao2019diffusive}. Furthermore, we assume that TX can only release the ICM, when it gets back to its previous position after the collision. These assumptions are validated through numerically and SPBS in Fig. \ref{fig:1AA} and Fig. \ref{fig3}.
\\
\underline{\textbf{Numerical Analysis of FHTD and Hitting Probability:}}
In Fig. \ref{fig:1AA}, the PDF $f_{R_k}(r_k)$ is plotted as a function of random distance $r_k$.
It can be observed that, at a given initial distance, i.e, at $r_{0} = 10\,\mu$m, for the higher mobility, i.e, at $D_{tx} = D_{rx} = 5\,\mu$m$^2$/s, the probability that the TX and RX remain on their initial positions is less as compared to the case, when the mobility of TX and RX is lesser i.e., for $D_{tx} = D_{rx} = 1\,\mu$m$^2$/s.
Moreover, it can also be observed that, there is a non-zero probability that the distance between the TX and RX becomes zero.
\\
In Fig. \ref{fig3} derived FHTD is verified and analyzed using SPBS.
For SPBS, at the time-step between two successive samples $\Delta t_h = 0.5$\,s, the locations of each molecule, TX and RX, are identified.
The value of $D_m = 5\,\mu$m$^2$/s and $r_0 = 10$\,$\mu$m. The change in position ($\Delta z_{\phi}$) of molecule, TX and RX along the \textit{z}-direction in each time-step is modeled independently and is assumed to follow a Gaussian distribution~\cite{farsad2016comprehensive}, i.e., $\Delta z_{\phi}\sim\mathcal{N}(0,2 D_{\phi} (\Delta t_h)^{\alpha})$, where $\phi\in\{m,tx,rx\}$. The distance between TX and RX is obtained for each time-step. After that, the FHTD values are obtained by averaging over all the random distances between TX and RX. 
In Fig. \ref{fig3}, it is observed that for all the diffusion phenomena, when both TX and RX are mobile, the value of hitting probability is lower as compared to the case when either TX or RX is mobile.
Since, a closed-form expression for the hitting probability is not mathematically tractable. Therefore, we use numerical integration to find the values of the hitting probability of molecule at the RX as $F_{h}(t) = \int_{0}^{t} \overline{f_{T_h}(t_h;k)}~dt_h$. 
\\
Fig. \ref{fig:3} shows the hitting probability as a function of time, it can be observed that hitting probability is monotonically non-decreasing function with time. From this figure, it can be observed that for lower mobility of RX, i.e., at $D_{rx} = 0.5\times 10^{-12}$\,m$^2$/s, any increase in the value of anomalous diffusion exponent ($\alpha$) leads to increase in the values of hitting probability at a given time. Whereas, for the higher mobility case, i.e., at  $D_{rx} = 50\times 10^{-12}$\,m$^2$/s, the value of hitting probability at a given time decreases with an increase in the value of anomalous diffusion exponent. This happens because, for higher values of $\alpha$, the RX also diffuses fast in the channel, which leads to a reduction in RXs' hitting probability.
\section{Communication Channel Performance}
\underline{\textbf{Formulation of Achievable Information Rate:}}
To reduce the mathematical cumbersomeness, TX and RX are considered to be perfectly synchronized \cite{huang2020clock}. Further, we assume that each molecule's motion is independent, and the ICM independently reach the RX.
We use timing modulation to transmit binary symbols from TX to RX in a binary channel with channel-use duration $T_u$. In this method, information is encoded based on the release time of a molecule. Note that, in order to reduce the effect of interference, the channel-use interval ($T_u$) can be chosen large enough, such that most of the ICM emitted are absorbed in the same channel-use interval, and are not received at the RX in other communication intervals.
Let $T_{r}$ ($r\in\{0,1\}$) be the release time for an information molecule for the transmission of binary symbol $X$ as either 0 or 1. However, due to the propagation delay, a molecule may not reach the RX in the expected time duration. The time at which RX absorbs the molecule is given as $T_a = T_r + T_h$.
Thus, there are three possible outcomes at the RX, corresponding to the arrival time $T_a$ of a molecule. The three states at the receiver denoted as $Y\in\{0,1,\epsilon\}$. Output state $Y$ is detected as either 0 or 1 if the molecule arrive in the assumed channel-use duration; otherwise, the receiver is in the erasure state,  i.e., $Y = \epsilon$ if the molecule does not reach in the given time-slot.
Let $p_0$ and $p_1$ be the probabilities that the molecule emitted at the time instances $T_0$ and $T_1$ are correctly received as 0 and 1, respectively, in the same channel-use duration $T_u$. Furthermore, $\epsilon_0$ and $\epsilon_1$ are the erasure probabilities corresponding to binary emissions 0 and 1, respectively.
If we consider $T_0 = 0$, the transition probabilities can be defined as
\vspace{-0.3cm}
\begin{align}
&p_{0} = F_{h}(\eta),~p_{1} = F_{h}(T_u-T_1)- F_{h}(\eta-T_1), \label{eq:1A}\\
&\epsilon_{0}{=}F_{h}(\infty){-}F_{h}(T_u),
\epsilon_{1}{=}F_{h}(\infty{-}T_1){-}F_{h}(T_u{-}T_1),\label{eq:1B}
\end{align}
\vspace{-0.1cm}
where $\eta$ denotes the decision threshold at the RX. Note that, in the considered timing modulation scheme, each time a transmitted molecule may not reach at the receiver, within the interval 0 to $\eta$ for bit 0, or within the interval $\eta$ to $T_u$ for bit 1. 
Considering the channel-use duration is long enough, the AIR for the equivalent timing channel with mutual information $I(T_r;T_a)$, corresponding to the release and arrival times of a molecule between TX and RX can be formulated as
\begin{align}
C &\approx\underset{\beta}{\max}~I(T_r;T_a) = \underset{\beta}{\max}~(H(T_a)-H(T_a|T_r)),
\end{align}
where $\beta$ denotes the input distribution over which the mutual information is maximized. The terms $H(\cdot)$ and $H(\cdot|\cdot)$ denote the binary entropy and binary conditional entropy functions, respectively. Here $T_{a}$ denotes the time of absorbing the molecule and it corresponds to the three possible outcomes, i.e., $Y\in\{0,1,\epsilon\}$ at the RX.
Now, we expand the terms $H(T_a)$ and $H(T_{a}|T_{r})$ as follows
\begin{align}
H(T_a) = -\underset{T_a\in\{0,1,\epsilon\}}{\sum}\text{Pr}(T_a)\log_2(\text{Pr}(T_a)),\label{eq:123}
\end{align}
where \text{Pr}($\cdot$) denotes the probability function,
and $H(T_a|T_r)  = \text{Pr}(T_r  = 0)H(T_a|T_r=0)+\text{Pr}(T_r = 1)H(T_a|T_r=1),$
here $\text{Pr}(T_r = 0) = \beta$ and $\text{Pr}(T_r = 1) = 1-\beta$. The probabilities in \eqref{eq:123} are defined as
\begin{align}
&\text{Pr}(T_a = 0) = \beta p_{0} + (1-\beta)(1-p_{1}-\epsilon_{1}),\\
&\text{Pr}(T_a = 1) = (1-\beta) p_{1} + \beta(1-p_{0}-\epsilon_{0}),\\
&\text{Pr}(T_a = \epsilon) = (1-\beta) \epsilon_{1} + \beta \epsilon_{0},
\end{align}
where $p_{0}$, $p_{1}$, $\epsilon_0$, and $\epsilon_1$ are defined in \eqref{eq:1A}-\eqref{eq:1B}.

\begin{figure}
  \centering
  \includegraphics[width=3.5in, height = 1.8in]{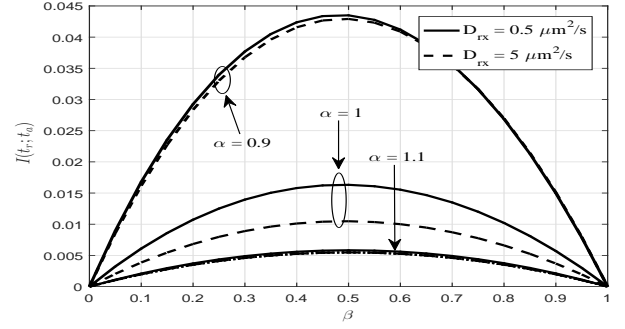}
  \vspace{-0.6cm}
  \caption{Mutual information as a function of $\beta$.}
  \vspace{-0.6cm}\label{fig:4}
\end{figure}

\underline{\textbf{Numerical Analysis of AIR:}} The mutual information as a function of input distribution $\beta$ is shown in Fig. \ref{fig:4}. For this figure, the system parameters are considered as follows: $D_{m}=5\,\mu$m$^2$/s, $r_{0}= 10\,\mu$m, $k=1$, $D_{tx} = 0$, $T_u = 40$\,s, and decision threshold $\eta$ is chosen as mean value of first hitting time. The value of $\eta$ is obtained as $\eta = \int_{0}^{T_s}t_h \overline{f_{T_h}(t_h;k)}~dt_h$ over the sampling interval $T_s = 1$\,s. 
In Fig. \ref{fig:4}, it can be shown that for the given system parameters, in super-diffusion case, i.e., at $\alpha = 1.1$, the AIR is lowest compared to the other two diffusion phenomena (normal-diffusion and sub-diffusion). This is due to the fact that for the super-diffusion phenomenon, RX diffuses more rapidly from its original position, which results in a lower hitting probability of molecule; consequently, it reduces the AIR.  In all the diffusion phenomena, the AIR further reduces as the mobility of TX increases from $0.5\,\mu$m$^{2}$/s to $5\,\mu$m$^{2}$/s.
This is due to the fact that for the high mobility, the hitting probability of molecule at the RX in a given channel-use duration reduces.
\vspace{-.4cm}
\section{Conclusion}
In this work, a one-dimensional anomalous-diffusive MC  channel, wherein the transmitter and receiver can move anomalously, is considered. To model the anomalous diffusion of ICM, transmitter, and receiver, the concept of time-scaled Brownian motion is exploited.
A novel closed-form expression for the FHTD is derived.
Further, the channel between transmitter and receiver is modeled as binary-erasure-channel.
Moreover, timing modulation is incorporated for the transmission of binary symbol, and the considered binary erasure channel is analyzed in terms of AIR.

\ifCLASSOPTIONcaptionsoff
  \newpage
\fi
\medskip

\vspace{-.8cm}
\bibliographystyle{IEEEtran}
\bibliography{IEEEabrv,Ref_list_Anomalous}

\end{document}